\documentclass[twocolumn,preprintnumbers,floatfix,showpacs,prb,superscriptaddress]{revtex4}
\usepackage{graphicx}
\usepackage{dcolumn} 
\usepackage{bm}
\usepackage{epsfig}
\usepackage{longtable}
\begin{document} 

\title{Electronic structure of CrN: A comparison between different exchange correlation potentials}

\author{A.~S. Botana}
\email[Corresponding author. Email address: ]{antia.sanchez@usc.es}
\affiliation{Departamento de F\'{i}sica Aplicada,
  Universidade de Santiago de Compostela, E-15782 Campus Sur s/n,
 Santiago de Compostela, Spain}
 \affiliation{Instituto de Investigaci\'{o}ns Tecnol\'{o}xicas,
  Universidade de Santiago de Compostela, E-15782 Campus Sur s/n,
  Santiago de Compostela, Spain}  
\author{F. Tran}
\affiliation{Institute of Materials Chemistry, Vienna University of Technology, Getreidemarkt 9/165-TC, A-1060 Vienna, Austria}
\author{V. Pardo}
\affiliation{Departamento de F\'{i}sica Aplicada,
  Universidade de Santiago de Compostela, E-15782 Campus Sur s/n,
  Santiago de Compostela, Spain}
\affiliation{Instituto de Investigaci\'{o}ns Tecnol\'{o}xicas,
  Universidade de Santiago de Compostela, E-15782 Campus Sur s/n,
  Santiago de Compostela, Spain}  
\author{D. Baldomir}
\affiliation{Departamento de F\'{i}sica Aplicada,
  Universidade de Santiago de Compostela, E-15782 Campus Sur s/n,
  Santiago de Compostela, Spain}
\affiliation{Instituto de Investigaci\'{o}ns Tecnol\'{o}xicas,
  Universidade de Santiago de Compostela, E-15782 Campus Sur s/n,
  Santiago de Compostela, Spain}  
\author{P. Blaha}
\affiliation{Institute of Materials Chemistry, Vienna University of Technology, Getreidemarkt 9/165-TC, A-1060 Vienna, Austria}

\date{\today}

\begin{abstract}

We report a series of electronic structure calculations for CrN using different exchange correlation potentials: PBE, LDA+$U$, the Tran-Blaha modified Becke-Johnson, and hybrid functionals. In every case, our calculations show that the onset of magnetism in CrN should be accompanied by a gap opening. The experimentally found antiferromagnetic order always leads to an insulating behavior. Our results give further evidence that the Tran-Blaha functional is very useful for treating the electronic structure of correlated semiconductors allowing a parameter free description of the system. Hybrid functionals are also well capable of describing the electronic structure of CrN. The analysis of the system is complemented with our calculations of the thermopower that are in agreement with the experimental data.

\end{abstract}

\maketitle

\section{Introduction}\label{intro}

Transition-metal nitrides (TMN) have drawn considerable attention due to their interesting physical properties: mechanical strength, ultrahardness, corrosion resistance, high melting points, good electrical and thermal conductivities, etc.\cite{review_nitrides} In addition, some TMN exhibit other properties such as superconductivity (arising in VN (Ref. \onlinecite{vn_zhao}) and NbN (Ref. \onlinecite{nbn_supercond, nbn_supercond_2})), a transition from a superconductor to a Cooper pair insulator in TiN (Ref. \onlinecite{tin_supercond}), or a magnetostructural phase transition (MPT) as a function of temperature in CrN (Ref. \onlinecite{corliss}).

\begin{figure}
\includegraphics[width=\columnwidth,draft=false]{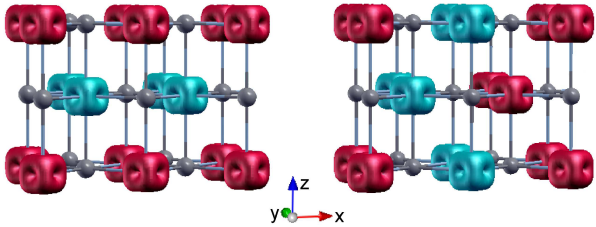}
\caption{(Color online) Three dimensional plot showing the difference between the spin-up and spin-down (different color) electron
densities in the AFM1 (left panel) and AFM2 (right panel) phases. The typical d$^3$(t$_{2g}^3$) electronic configuration of each Cr$^{3+}$ cation and no magnetic polarization in neighboring ligands can be observed.}\label{AFM}
\end{figure}

This MPT in CrN has attracted considerable attention and controversial results have been reported. Some studies have shown that this transition is linked to a discontinuity in the resistivity versus temperature curve. \cite{crn_constantin,crn_quintela_apl,crn_quintela_prb,bhobe} It has been characterized as a first-order MPT from a high-temperature paramagnetic (PM), cubic NaCl-type phase to a low-temperature antiferromagnetic (AFM) orthorhombic \textit{Pnma} one at a N\'eel temperature $T_N\sim$ 273-286 K.\cite{corliss} The low-$T$ AFM configuration is quite unique consisting of double ferromagnetic (FM) layers stacked antiferromagnetically along the [110] direction (see AFM2 in the right panel of Fig. \ref{AFM}).
From neutron powder diffraction experiments, a magnetic moment of 2.36~$\mu_{B}$ per Cr atom was obtained  by Corliss \textit{et al}.,\cite{corliss} while in Ref. \onlinecite{IbbersonPB92} a value of 3.17~$\mu_{B}$ was measured.
This peculiar magnetic structure has been shown to be linked to the structural distortion across $T_N$.\cite{filippetti_prb, filippetti} However, other works show the suppression of the phase transition due to epitaxial constraints when grown in films with the system remaining in the cubic phase at all temperatures. \cite{gall_jap_1, inumaru_prb,gall_epitaxial_suppression}

The main focus of controversy has been the understanding of CrN electrical behavior. On the one hand, the values of the resistivity at room temperature vary over several orders of magnitude (from m$\Omega$cm to k$\Omega$cm) in different works. \cite{anderson, bhobe, crn_constantin, gall_jap_1, gall_jap_2, gall_vrh, gall_epitaxial_suppression, crn_quintela_apl, tsuchiya, crn_quintela_prb} On the other, the nature of the transition in terms of the electrical conduction behavior has had a wide range of interpretations in the literature:

\textit{ Metal to metal.} In the work of Tsuchiya \textit{et al.},\cite{tsuchiya} polycrystalline CrN$_{1+x}$ films with 0$\leq$ x$\leq$ 0.2 have been grown by reactive sputtering. Nearly stoichiometric films show an AFM first order phase transition at $T_N\sim260$ K. Resistivity measurements for these samples indicated metallic behavior across $T_N$ with a steep decrease of the resistivity below the transition. The same behavior was found by Browne \textit {et al.} \cite{browne} in CrN powder. In the study of Inumaru \textit{et al.},\cite{inumaru_prb} CrN films were epitaxially grown on $\alpha$-Al$_2$O$_3$ (0001) and MgO (001) substrates by pulsed laser deposition. The films grown on
$\alpha$-Al$_2$O$_3$ (0001) with its (111) planes parallel to the substrate show no structural transition whereas CrN(001) films grown on MgO (001) show a transition to an AFM state at about 260 K. Both films exhibit metallic $T$-dependence of the resistivity, but a drop in its values below the N\'eel temperature is observed for the samples showing a MPT. 

\textit{ Metal to semiconductor.} Constantin \textit{et al.}\cite{crn_constantin} grew CrN(001) thin films on MgO(001) using molecular beam epitaxy. The films show a MPT at 285 K, with activated semiconducting behavior above $T_N$ and a band gap of 0.07 eV. Below the transition temperature, the films are metallic. The polycrystalline samples of CrN prepared by Bhobe \textit{et al.}\cite{bhobe} show a MPT at 286 K. Above $T_N$, CrN exhibits activated semiconducting behavior (with a gap of 0.07 eV) becoming a disordered metal below $T_N$. Evidences of strong electron-electron correlations were found, determining an on-site Coulomb energy ($U$) of 4.5 eV by resonant photoemission spectroscopy (PES). In Ref. \onlinecite{gall_epitaxial_suppression} both single and polycrystalline CrN layers have been grown by reactive sputtering deposition on both MgO and quartz substrates. Polycrystalline CrN did show a disorder induced metal-insulator transition to an orthorhombic phase across $T_N\sim280$ K. 

\textit{ Semiconductor to semiconductor.} In the above mentioned work of Tsuchiya \textit{et al.},\cite{tsuchiya} the non-stoichiometric films did not show a MPT. 	The resistivity decreases monotonically with $T$ suggesting a semiconducting (hopping-like) behavior of the samples. CrN powder samples synthesized by Herle \textit{et al.}\cite{herle} show a transition from a PM to an AFM phase at 280 K. Activated semiconducting behavior with a band gap of 0.09 eV was obtained from resistivity vs. $T$ curves. Quintela \textit {et al.}\cite{crn_quintela_prb} studied polycrystalline samples of Cr$_{1-x}$V$_x$N. Stoichiometric CrN shows a MPT at $T_N\sim286$ K and semiconducting behavior in the whole $T$-range. Above $T_N$, the resistivity vs. $T$ curve shows activated semiconducting behavior with a band gap of 150 meV. Below the transition, the changes in the chemical bond linked to the AFM order, give as a result a non-activated semiconducting behavior. The thermoelectric power of stoichiometric CrN shows a linear $T$-dependence as expected for metals or highly degenerate semiconductors. This supports the idea of an electronic state neither thermally activated nor fully itinerant in the AFM phase. Further evidence of this idea is given by the drastic reduction of the bulk modulus due to the  bond softening and charge delocalization into the shorter and less strong Cr-Cr bonds in the AFM phase.\cite{fran_nat_mat} By ultrahigh vacuum magnetron sputtering, Gall \textit{et al.} \cite{gall_jap_2, gall_jap_1, gall_vrh, gall_vibrational_modes, gall_epitaxial_suppression} grew single crystalline CrN (001) films on a MgO(001) substrate. The absence of any discontinuity in the resistivity vs. $T$ curves suggests that epitaxial constraints suppress the transition to an AFM phase with the film being cubic PM at all T. The resistivity decreases with increasing temperature indicating semiconducting behavior over the whole T-range. From optical absorption measurements, an optical band gap of 0.7 eV was obtained.\cite{gall_jap_1, anderson} An indirect band gap of about 0.19 eV was also estimated \cite{gall_vibrational_modes} from optical measurements suggesting CrN is a Mott-Hubbard type insulator. In addition, the samples show semiconducting behavior in a variable range hopping regime below 120 K.\cite{gall_vrh, gall_epitaxial_suppression} 

As can be seen, there is a big discussion and disagreement between different experimental works. There are also some theoretical studies on CrN. Calculations within the local density approximation (LDA) have been performed on CrN by Filippetti \textit{et al.}\cite{filippetti_prb, filippetti} For the distorted othorhombic structure, they found as ground state the experimentally observed AFM2 configuration. For the cubic structure, the ground state is the so-called AFM1, similar to the AFM2, but with the spins antialigned every layer instead of every two layers
(left panel of Fig. \ref{AFM}). This can be understood in terms of the magnetic stress: the AFM2 phase is under more stress with respect to the AFM1. This is due to the spin ordering asymmetry of the AFM2 phase on the (001) plane where each Cr has two spin-antiparallel and two spin-parallel nearest neighbors. Considering that bonds between the antiferromagnetically coupled neighbors are under tensile stress, while bonds between ferromagnetically coupled atoms are under compressive stress, in the AFM2 magnetic configuration, stress can be relieved by an orthorhombic distortion. In the AFM1 structure, it cannot be relieved because all the four-nearest neighbors are FM coupled. Hence, in order to relieve stress and reduce the total energy, the system undergoes a shear distortion from cubic to orthorhombic. 
LDA calculations show a metallic behavior for both AFM phases (weak metal since the density of states (DOS) at the Fermi level is small). From studies of TM oxides and other correlated compounds it is known that LDA underestimates the Coulomb effects on the narrow $d$ bands, predicting an unrealistic metallic state in systems that actually have a band gap. Hence, further electronic structure calculations in CrN within the LDA+$U$\cite{sic} method were performed by Herwadkar \textit{et al.}\cite{herd} If the on-site Coulomb repulsion ($U$) is considered, a band gap already opens for a value of $U=3$ eV in the magnetic ground state, suggesting that CrN may be close to a charge-transfer insulator.

To shed light on the existing controversies about CrN, we present the results of renewed \textit{ab initio} calculations using different computational schemes to treat the moderately correlated $d$ electrons of CrN and describe properly its electronic structure and transport properties.

\section{COMPUTATIONAL DETAILS}

Our electronic structure calculations were performed with the WIEN2k code,\cite{wien2k,wien} based on density functional theory (DFT)\cite{dft,lda,dft_2} utilizing the augmented plane wave plus local orbitals method (APW+lo).\cite{sjo} For the calculations of the transport properties we used the BoltzTraP code,\cite{boltztrap} a semiclassical approach based on the Boltzmann transport theory.

Most of the DFT-based calculations in solids have been done using two (semi)local approaches for the exchange correlation energy and potential: local density approximation (LDA)\cite{lda} or generalized gradient approximation (GGA).\cite{gga_1} Although they give useful total energies and structural parameters, both of them are known to underestimate the band gap of most semiconductors and insulators. 

There are several functionals that can provide more accurate values for band gaps. On the one hand, staying inside the true Kohn-Sham (KS) framework (multiplicative potential), the newly developed semilocal potentials (such as the one developed by Tran and Blaha based on a modification of the Becke-Johnson\cite{BJ} potential) can give good results.\cite{mbj, singh_mbj} But on the other hand, nonmultiplicative potentials can also be used. These methods belong to the so-called generalized KS framework\cite{g_ks_framework} and most of them mix DFT and Hartree-Fock (HF) theories. The most used ones are the LDA+$U$ \cite{sic} approach and hybrid methods.\cite{hybrid}

The LDA+$U$ scheme improves over GGA or LDA in the study of systems containing correlated electrons by introducing the on-site Coulomb repulsion $U$ applied to localized electrons (e.g $3d$ or $4f$). We have performed calculations on CrN within the LDA+$U$ method (using the fully localized version for the double-counting correction\cite{CzyzykPRB94}),
taking $U$ in a reasonable range for this type of $3d$ electron systems (from 2 to 5 eV) comparable to the values obtained by PES experiments. The exchange parameter $J$ was set to zero.

In hybrid methods a certain amount of semilocal exchange is replaced by HF exchange with the correlation remaining purely semilocal. They usually lead to improved band structures in good agreement with experiments as shown for several semiconductors and insulators or even strongly correlated electron systems.\cite{hybrids_1,hybrids_2, hybrids_3} We have applied one of the most common hybrid functionals (PBE0)\cite{PBE0, PBE0_2} to the study of the electronic structure of CrN. However, traditional hybrid functionals are computationally very demanding due to the long-range nature of the HF exchange in solids.
A way to reduce this problem of convergence is to consider only the short-range part of the HF exchange
as proposed by Bylander and Kleinman.\cite{BylanderPRB90} This scheme has been used by
Heyd, Scuseria, and Ernzerhof\cite{hse_1, hse_2} who proposed a screened hybrid functional based
on PBE0. We have also performed calculations using a screened hybrid functional as implemented in the WIEN2k code with the HF exchange screened by means of the Yukawa potential (so called, YS-PBE0).\cite{prb_hybrids_yukawa_lapw}

The modified Becke-Johnson exchange potential (a local approximation to an atomic exact-exchange potential and a screening term) + LDA-correlation (from hereon TB-mBJLDA) allows the calculation of band gaps with an accuracy similar to the much more expensive $GW$ or hybrid methods.\cite{mbj,KollerPRB11,singh_mbj} We have also studied the electronic structure of CrN by using the TB-mBJLDA potential which does not contain any system-dependent parameter.

The calculations were done with $R_{MT}K_{max}=7$ (the product of the smallest of the atomic sphere radii $R_{MT}$
and the plane wave cutoff parameter $K_{max}$), which determines the size of the basis set.
The chosen $R_{MT}$ were 2.06 an 1.82 for Cr and N, respectively.
The PBE, LDA+$U$, and TB-mBJLDA calculations were done with a $7\times13\times9$ k-mesh
for the integrations over the Brillouin zone. Since calculations with hybrid functionals are much more costly,
a less dense k-mesh of $4\times8\times6$ was used for the YS-PBE0 and PBE0 functionals. Convergence for the total energies presented is achieved with that k-mesh. The transport properties are very sensitive to the Brillouin zone sampling, therefore
an even denser grid ($19\times37\times27$) was used to obtain convergence.

As mentioned in Sec. \ref{intro}, the transition at $T_{N}$ in CrN from the PM to the AFM phase
is accompanied by a distortion of the rocksalt structure. The shear distortion consists of a contraction of the angle from the cubic (90$^\circ$) to 88.23$^\circ$ (see Ref. \onlinecite{filippetti_prb} for a more detailed description).
Calculations were done both, with and without distortion at
the experimental lattice constants (Ref. \onlinecite{corliss}). In order to accomodate the AFM2
phase, a unit cell comprising four formula units has been used. The calculations
for the other magnetic phases were also performed with this unit cell.

\section{RESULTS}

\begin{table*}
\caption{Relative energy (in meV/formula unit), fundamental band gap (in eV),
and Cr atomic moment (in $\mu_{B}$) of CrN for the different methods used
in this work. It was not possible to converge the calculations with the hybrid
functionals for the metallic NM phase.\label{table1}}
\begin{ruledtabular}
\begin{tabular}{lcccccccc}
\multicolumn{1}{l}{} &
\multicolumn{4}{c}{orthorhombic} &
\multicolumn{4}{c}{cubic} \\
\cline{2-5}\cline{6-9}
\multicolumn{1}{l}{} &
\multicolumn{1}{c}{AFM2} &
\multicolumn{1}{c}{AFM1} &
\multicolumn{1}{c}{FM} &
\multicolumn{1}{c}{NM} &
\multicolumn{1}{c}{AFM2} &
\multicolumn{1}{c}{AFM1} &
\multicolumn{1}{c}{FM} &
\multicolumn{1}{c}{NM} \\
\hline
Relative energy \\
PBE                             & 0 & 20 & 211 &  504 & 17 & 13 & 208 &  507 \\
LDA+$U$ ($U_{\text{eff}}=2$ eV) & 0 & 42 & 215 & 1068 & 14 & 35 & 209 & 1074 \\
LDA+$U$ ($U_{\text{eff}}=3$ eV) & 0 & 49 & 202 & 1540 & 11 & 41 & 194 & 1548 \\
LDA+$U$ ($U_{\text{eff}}=4$ eV) & 0 & 50 & 180 & 2028 &  8 & 41 & 170 & 2049 \\
LDA+$U$ ($U_{\text{eff}}=5$ eV) & 0 & 51 & 156 & 2405 &  5 & 41 & 146 & 2456 \\
YS-PBE0 ($\alpha=0.1$)          & 0 & 34 & 270 &      & 12 & 25 & 261 &      \\
PBE0 ($\alpha=0.1$)             & 0 & 36 & 279 &      & 12 & 26 & 270 &      \\
YS-PBE0 ($\alpha=0.25$)         & 0 & 34 & 232 &      &  7 & 23 & 220 &      \\
PBE0 ($\alpha=0.25$)            & 0 & 35 & 229 &      &  7 & 24 & 218 &      \\
\hline
Band gap \\
PBE                             & 0    & 0    & 0    & 0    & 0    & 0    & 0    & 0 \\
LDA+$U$ ($U_{\text{eff}}=2$ eV) & 0    & 0    & 0    & 0    & 0    & 0    & 0    & 0 \\
LDA+$U$ ($U_{\text{eff}}=3$ eV) & 0.30 & 0    & 0    & 0    & 0.30 & 0    & 0    & 0 \\
LDA+$U$ ($U_{\text{eff}}=4$ eV) & 0.68 & 0.17 & 0    & 0    & 0.66 & 0.20 & 0    & 0 \\
LDA+$U$ ($U_{\text{eff}}=5$ eV) & 0.91 & 0.51 & 0    & 0    & 0.89 & 0.56 & 0    & 0 \\
TB-mBJLDA                       & 0.79 & 0.22 & 0    & 0    & 0.80 & 0.27 & 0    & 0 \\
YS-PBE0 ($\alpha=0.1$)          & 0.20 & 0    & 0    &      & 0.20 & 0    & 0    &   \\
PBE0 ($\alpha=0.1$)             & 0.48 & 0    & 0    &      & 0.47 & 0    & 0    &   \\
YS-PBE0 ($\alpha=0.25$)         & 1.45 & 0.84 & 0    &      & 1.44 & 0.95 & 0    &   \\
PBE0 ($\alpha=0.25$)            & 2.13 & 1.53 & 0.43 &      & 2.11 & 1.66 & 0.47 &   \\
\hline
Cr spin magnetic moment \\
PBE                             & 2.32 & 2.23 & 2.28 & 0 & 2.33 & 2.23 & 2.30 & 0 \\
LDA+$U$ ($U_{\text{eff}}=2$ eV) & 2.51 & 2.43 & 2.61 & 0 & 2.51 & 2.43 & 2.61 & 0 \\
LDA+$U$ ($U_{\text{eff}}=3$ eV) & 2.59 & 2.52 & 2.70 & 0 & 2.59 & 2.52 & 2.70 & 0 \\
LDA+$U$ ($U_{\text{eff}}=4$ eV) & 2.65 & 2.58 & 2.74 & 0 & 2.65 & 2.58 & 2.74 & 0 \\
LDA+$U$ ($U_{\text{eff}}=5$ eV) & 2.69 & 2.62 & 2.76 & 0 & 2.69 & 2.62 & 2.76 & 0 \\
TB-mBJLDA                       & 2.52 & 2.48 & 2.64 & 0 & 2.52 & 2.49 & 2.64 & 0 \\
YS-PBE0 ($\alpha=0.1$)          & 2.49 & 2.42 & 2.62 &   & 2.49 & 2.42 & 2.63 &   \\
PBE0 ($\alpha=0.1$)             & 2.49 & 2.42 & 2.64 &   & 2.50 & 2.43 & 2.65 &   \\
YS-PBE0 ($\alpha=0.25$)         & 2.59 & 2.54 & 2.67 &   & 2.59 & 2.54 & 2.67 &   \\
PBE0 ($\alpha=0.25$)            & 2.59 & 2.54 & 2.67 &   & 2.60 & 2.54 & 2.67 &   \\
\end{tabular}
\end{ruledtabular}
\end{table*}

Since the compound is a moderately correlated TMN, a crude estimate of the possible electronic configuration of the material can be obtained from an ionic model. Taking the usual valency for N, the average valence for the octahedrally coordinated Cr is +3. In a high spin state ($S= 3/2$, magnetic moment of 3 $\mu_{B}$), the threefold-degenerate majority-spin channel $t^{\uparrow}_{2g}$ is completely filled while the minority $t^{\downarrow}_{2g}$ remains empty with the $e_g$ states for both spin channels also completely empty above the Fermi level. This electronic configuration can be observed in Fig.\ref{AFM}. Depending on the balance between crystal-field splitting ($\Delta_{CF}$) and Hund's rule coupling strenght ($J_H$), the first unoccupied bands could be $e^{\uparrow}_{g}$ ($J_H>\Delta_{CF}$) or $t^{\downarrow}_{2g}$ ($J_H<\Delta_{CF}$). Significant contribution of the N $p$ bands below the Fermi level is expected due to hybridization of these states with the Cr $d$ bands.

\begin{figure*}
\includegraphics[width=16.50cm,draft=false]{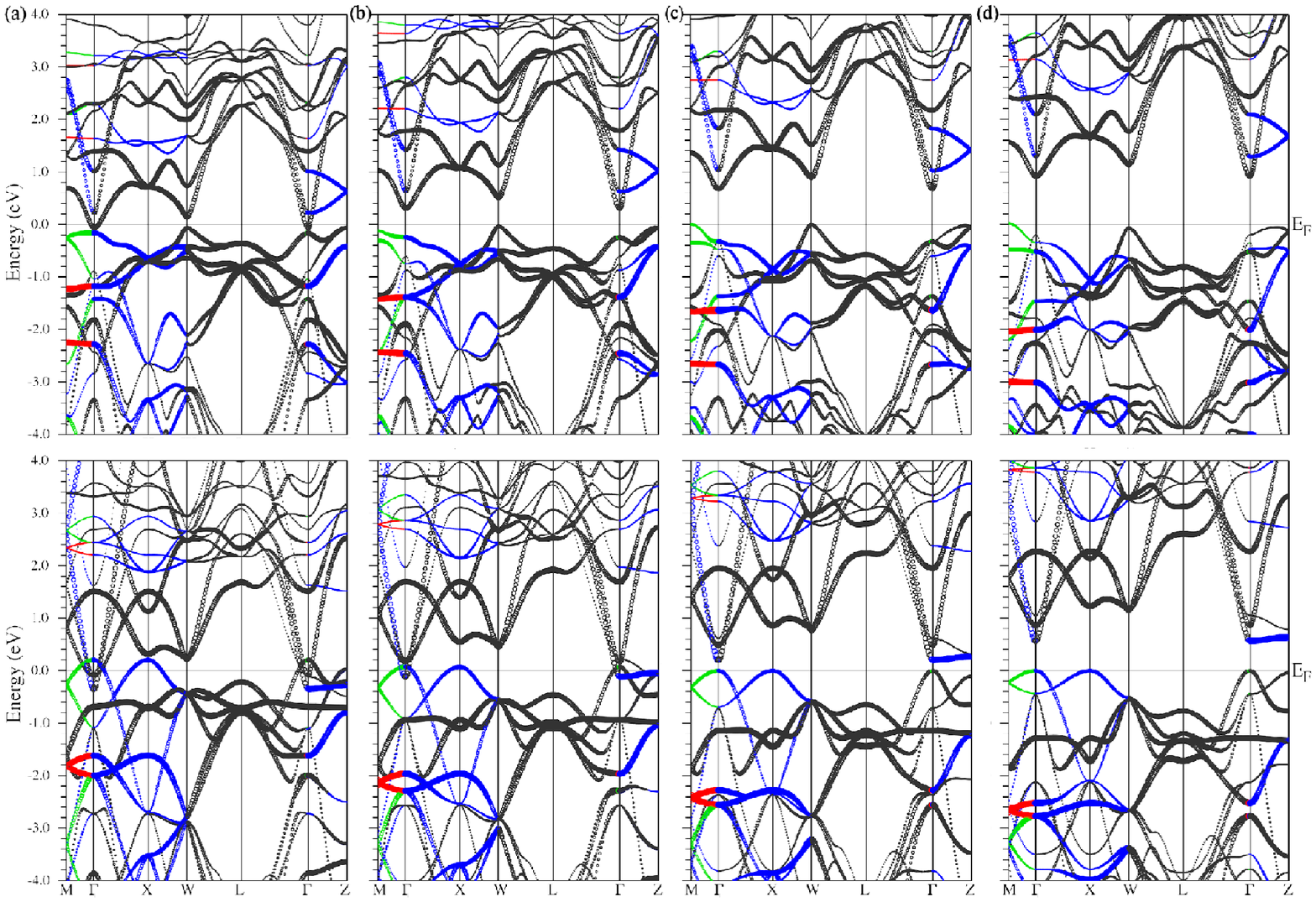}
\caption{(Color online) Band structures with band character plot (Cr $d$ highlighted) of the orthorhombic AFM2 phase (upper panels) and cubic AFM1 (lower panels) of  CrN obtained from the LDA+$U$ method with (a) $U=2$ eV, (b) $U=3$ eV, (c) $U=4$ eV, and (d) $U=5$ eV. The colors indicate the irreducible representation of the eigenvalues.}\label{bands}
\end{figure*}

In order to analyze in more detail the electronic structure of the compound, we have performed calculations with various methods: PBE (which is of the GGA form), LDA+$U$ with several values of $U$ (2-5 eV), TB-mBJLDA, and hybrid functionals (screened and unscreened). Table \ref{table1} shows the relative energies, band gaps, and magnetic moment on the Cr cation obtained for the AFM1, AFM2, FM and non-magnetic (NM) configurations. 

Looking at Table \ref{table1}, it can be seen that the orthorhombic AFM2 phase is favored overall, independent on the functional. For the cubic phase, this magnetic arrangement is also the more favored for the LDA+$U$ and hybrid functionals. The other phases are clearly less stable.
For instance, with LDA+$U$
AFM1 is disfavored by 35-41 meV/formula unit, FM by 146-209 meV/formula unit, and NM by
1074-2456 meV/ formula unit. Hybrid functionals retain the relative energies between the AFM2 and AFM1 phases and make the FM phase even more unstable with respect to the ground state AFM2. It has to be remarked that for PBE within the cubic phase (as it occured with LDA in the calculations carried out by Filippetti \textit{et al.}\cite{filippetti_prb, filippetti}), the AFM1 phase is more favored than the AFM2 although by just 4 meV/formula unit. An important point to note is that if the AFM1 is kept, no orthorhombic distortion would be predicted, independently of the functional used (i.e., the cubic AFM1 phase is always more stable than the orthorhombic AFM1 one). This is a consequence of the magnetostriction/stress relief issues described above and already found by Filipetti \textit{et al.} using LDA.\cite{filippetti_prb, filippetti} 

Concerning the band gap values, both the NM and FM configurations remain metallic for all the functionals except PBE0 with $\alpha=0.25$, which gives a band gap around 0.45 eV for FM. In LDA+$U$ calculations with $U=3$ eV (both for cubic -undistorted- and orthorhombic -distorted- structures), a gap of 0.3 eV already opens for the AFM2 configuration while the AFM1 remains metallic for this particular value of $U$ (a gap of about 0.2 eV opens for the AFM1 ordering at $U=4$ eV). If the value of $U$ is increased up to 5 eV, the value of the band gap is 0.9 eV for the AFM2 configuration and about 0.5 eV for AFM1. Hence, from LDA+$U$ calculations it can be concluded that the opening of a band gap is easier for the AFM2 (ground state) phase. The same conclusions can be drawn from the TB-mBJLDA calculations: both AFM configurations are insulating with values of the band gap similar to those obtained within LDA+$U$ for $U=4$ eV (0.79 and 0.22 eV for the AFM2 and AFM1 distorted structure, respectively). For hybrid functionals, the values of the band gap are very dependent on the amount of HF exchange ($\alpha$) and on whether it is screened or not. For $\alpha=0.25$, PBE0 gives a band gap for the AFM2 distorted phase of about 2.1 eV and of 1.5 eV for AFM1, whereas the screened YS-PBE0 gives much lower values of the band gap (1.45 eV for AFM2 and 0.84 eV for AFM1). Furthermore, the reduction of $\alpha$ to 0.1, leads to a huge reduction of the gap values, making the AFM1 phase metallic and reducing significantly the AFM2 band gap.

Concerning the magnetic moments of Cr ions, the values for LDA+$U$ are in between the reported
experimental values of 2.36 $\mu_{B}$\cite{corliss} and 3.17 $\mu_{B}$\cite{IbbersonPB92}
and increase with $U$. They change from 2.51 to 2.69 $\mu_{B}$ for the AFM2 configuration and from 2.43 to 2.62 $\mu_{B}$ for the AFM1 when varying $U$ from 2 to 5 eV. Slightly larger values are obtained for the FM phase (from 2.61 to 2.76 $\mu_{B}$), reaching values close to the atomic limit for a $d^3$ cation in the high-spin state. This increase is due to the reduction of the $p$-$d$ hybridization caused by $U$. For TB-mBJLDA, the magnetic moments are close to the
values obtained from LDA+$U$ with low values of $U=2$-3 eV. Hybrid functionals (both screened and unscreened) give magnetic moments which are slightly larger (smaller)
than TB-mBJLDA for $\alpha=0.25$ ($\alpha=0.1$).

Below, a more detailed discussion of the electronic structure is given for each
functional separately.

\begin{figure}
\includegraphics[width=0.98\columnwidth,draft=false]{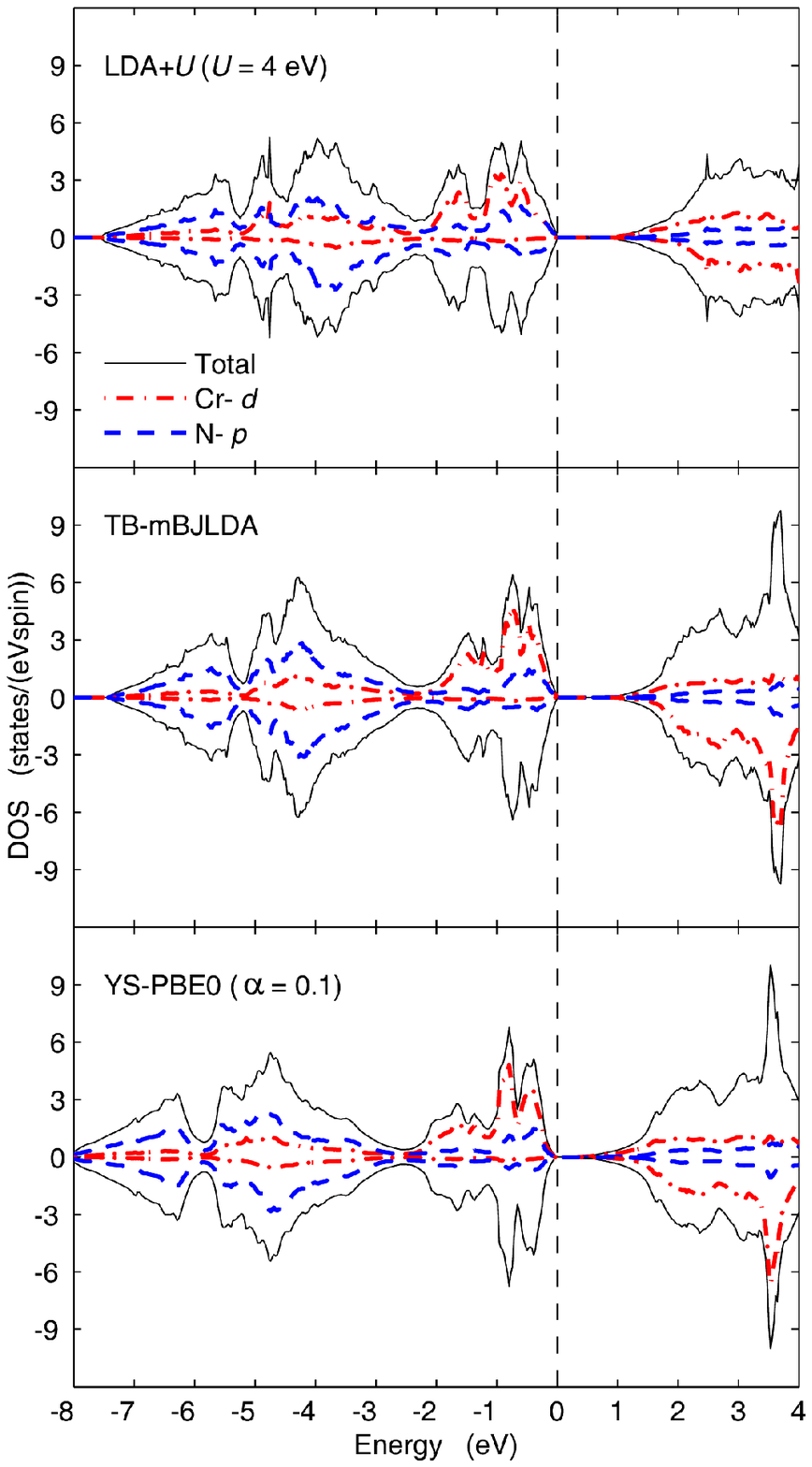}
\caption{\label{fignew}(Color online) Density of states of CrN
(orthorhombic AFM2 phase) calculated
with different methods. The Fermi energy is set at $E=0$ eV. The positive and negative values
on the $y$-axis are for spin-up and spin-down, respectively.}
\label{dos}
\end{figure}

\subsection{LDA+$U$ electronic structure analysis}

The evolution of the band structure with the $U$ value for both the cubic AFM1 and orthorhombic AFM2 phases can be seen in Fig. \ref{bands}. We use an orthorhombic Brillouin zone in all cases and the following k-path:
$M$($\pi/a$, 0, 0)-$\Gamma$(0, 0, 0)-$X$(0, $\pi/b$, 0)-$W$($\pi/a$, $3\pi/2b$, 0)-$L$(0, $\pi/b$, $\pi/c$)-$\Gamma$-$Z$(0, 0, $\pi/c$).
In Fig. \ref{dos}, the density of states (DOS) for the orthorhombic AFM2 phase (for $U= 4$ eV) is also shown.

For both magnetic configurations, the rough ionic electronic structure described above for the majority spin state of Cr atoms can be easily distinguished: the $t^{\uparrow}_{2g}$ manifold is filled
(in the range between $-2$ and 0 eV below the Fermi energy),
while only a small amount of (bonding) $e_g$ levels (in the range between $-6$ and $-3$ eV and
hybridizing with the N $p$ orbitals) are occupied. As can be seen in Fig. \ref{dos}, the band gap character is clearly $d$-$d$ ($t^{\uparrow}_{2g}$-$e^{\uparrow}_{g}$) although some rather small contribution of N $p$ states is also observed. Hence, CrN can be better described as a Mott-Hubbard insulator than as a charge-transfer insulator.\cite{ct_vs_mh} 

\begin{figure}
\includegraphics[width=\columnwidth,draft=false]{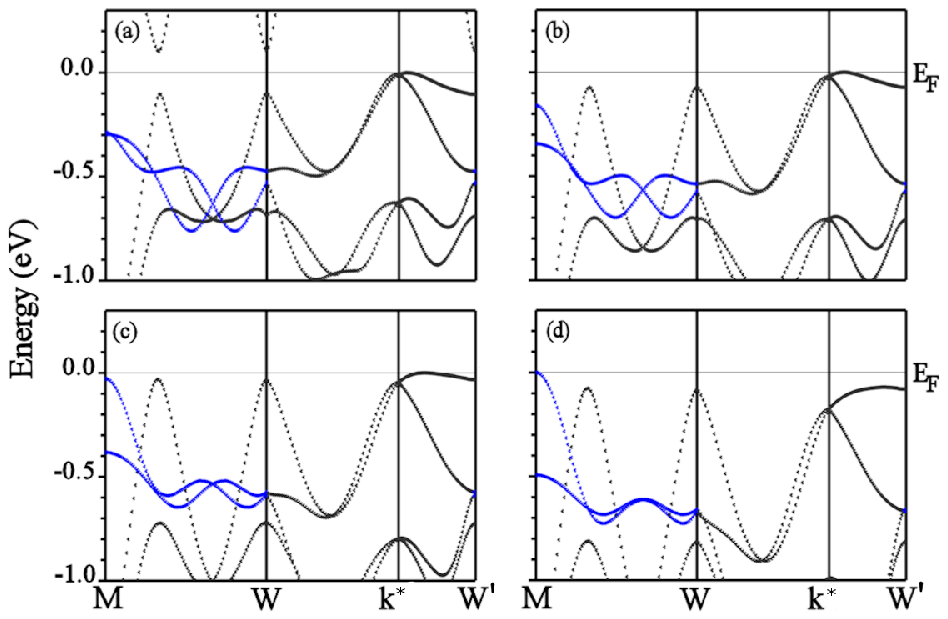}
\caption{(Color online) Band structures blown-up around E$_F$ for the orthorhombic AFM2 phase within LDA+$U$ [(a) $U=2$ eV, (b)
$U=3$ eV, (c) $U=4$ eV, and (d) $U=5$ eV] in a special k-path showing
the displacement of the VBM with the $U$ value.
The colors indicate the irreducible representation of the eigenvalues.}\label{bands_special}
\end{figure}

For the AFM2 configuration (upper panels of Fig. \ref{bands}), it can be seen how for the lowest $U$ value, the band structure remains metallic with a gap opening already for $U=3$ eV. The unoccupied bands move up gradually as U increases. This is particularly noticeable for the first unoccupied band at $\Gamma$ for $U=2$ eV and also the one that forms the small electron pocket. Also, as U increases, the bands right below the Fermi level move down in energy. For  larger $U$, the Cr $d$ character near $M$ is reduced since the Cr $t_{2g}^{\uparrow}$ states are shifted down in energy. The splitting between the first two bands right below the Fermi energy decreases at $\Gamma$ with increasing $U$, increases at $M$ and $W$ causing in principle a displacement of the top of the valence band from $W$ for lower $U$ values to $M$ for higher ones (for $U=4$ eV, almost the same eigenvalue is obtained at $M$ and $W$). Some features of the band structure are U-independent: the conduction band minimum is always at $\Gamma$ and the existence of degenerate bands at $L$ which is not accidental but symmetry required.

Another observation that can be made by looking at the changes in the DOS,
is a decrease of the distance between the main peaks of the occupied Cr $t_{2g}$
and N $p$ bands.
For $U=2$ eV, the distance is approximately 4 eV, while for $U=5$ eV it is reduced to
about 2.5 eV.

Previous band structure calculations \cite{herd} reported that when the gap
first opens at low $U$, the top of the valence band (VBM) is at $W$ moving
to $M$ as $U$ increases. Figure \ref{bands_special} shows the LDA+$U$ band
structure along the path $M$ ($\pi/a$, 0, 0)-$W$ ($\pi/a$, $3\pi/2b$, 0)-
$k.1$ ($\pi/a$, $3\pi/2b$, $\pi/c$)-$W'$ ($\pi/a$, $\pi/2b$, 0).
Our calculations show that when the gap opens (at $U\sim3$ eV), the VBM
is not at $W$ but displaced to a non-high symmetry point in the direction
from $k.1$ to $W'$, and then moving to $M$ at $U=5$ eV.

For the cubic AFM1 phase, a small gap of 0.20 eV is opened for $U=4$ eV becoming 0.56 eV for $U=5$ eV. We can see this evolution in the lower panels of Fig. \ref{bands}. In addition, numerous degenerate bands can be seen at $M$ and $W$ (with these degeneracies clearly lifted in the orthorhombic AFM2 phase).

\subsection{TB-mBJLDA electronic structure analysis}

As mentioned above, the band gap values obtained for TB-mBJLDA are rather similar to those obtained from LDA+$U$ with $U=4$ eV (Bhobe \textit{et al.}\cite{bhobe} determined an experimental value of $U$ of about 4.5 eV by PES). 

\begin{figure}
\includegraphics[width=\columnwidth,draft=false]{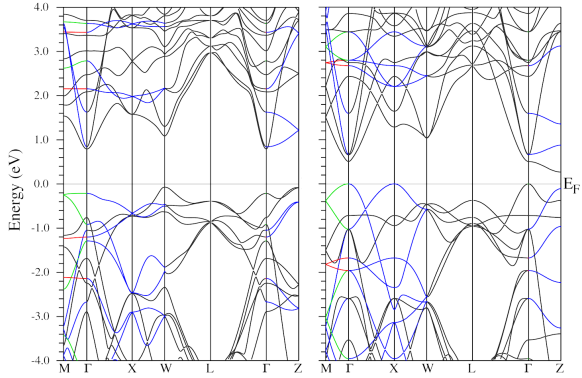}
\caption{(Color online) Band structures with TB-mBJLDA for the orthorhombic AFM2 (left panel) and cubic AFM1 (right panel) phase. The colors indicate the irreducible representation of the eigenvalues.}\label{bands_mbjlda}
\end{figure}

For the cubic AFM1 configuration, the band structure plot is shown in Fig. \ref{bands_mbjlda} (right panel). Below the Fermi level it is fairly different from the one obtained within LDA+$U$ (with low $U$). In addition, the first band above the Fermi level at $\Gamma$ in LDA+$U$ calculations lies lower in energy than with TB-mBJLDA, but it is not just a rigid band shift, some differences are noticeable particularly around the $\Gamma$ point. Actually, the bottom of the conduction band in TB-mBJLDA is no longer at $\Gamma$ but displaced to $Z$.  In addition, the band gap character changes from LDA+$U$ results being in this case a $t^{\uparrow}_{2g}$-$t^{\downarrow}_{2g}$ band gap.

For the orthorhombic AFM2 phase, the splitting at $\Gamma$ between the first two bands above the Fermi level is reduced in TB-mBJLDA (see left panel of Fig. \ref{bands_mbjlda}). The band structure obtained below the Fermi energy is very similar to the one obtained from LDA+$U$ with $U=2$ eV. In addition, as for LDA+$U$ calculations with the values of $U$ 3 to 4 eV, the VBM is not at $W$ but displaced to a non-high symmmetry point in the direction $k.1-W'$.
In Fig. \ref{dos} we can compare the DOS calculated from LDA+$U$ and TB-mBJLDA.
We can see that the N $p$ character in the Cr $t_{2g}^{\uparrow}$
band (from $-2.5$ to 0 eV) is smaller in TB-mBJLDA than in LDA+$U$ (and the opposite
is true for the Cr $e_{g}^{\uparrow}$ character in the N $p$ bands below -3 eV). This effect has to do with the different hybridizations obtained with the two schemes (enhanced with the LDA+U method). In addition,
the separation of N $p$ and Cr $t_{2g}^{\uparrow}$ bands is larger in TB-mBJLDA,
although the total valence band width is almost the same ($\sim7.5$ eV)

The optical transmission and reflection spectra of CrN are shown in Ref. \onlinecite{anderson}.
A band-gap-like optical transmission feature is found at 0.7 eV suggesting that CrN is a semiconductor with an indirect gap of approximately 0.7 eV and a $\Gamma$-valley gap greater than 0.7 eV. The values of the band gap (indirect) obtained from LDA+$U$ ($U=4$ eV) and TB-mBJLDA calculations for the AFM2 phase are very close to this 0.7 eV value (0.68 eV and 0.79 eV, respectively). A gap at $\Gamma$ of about 1 eV ($>0.7$ eV) is also found in our calculations. 

Thus, TB-mBJLDA is able to reproduce some of the effects on the electronic structure caused by the on-site Coulomb repulsion between Cr $d$ electrons without using a tunable parameter and being consistent with the results found for bulk CrN stoichiometric samples.

\subsection{Hybrid functionals electronic structure analysis}

\begin{figure}
\includegraphics[width=\columnwidth,draft=false]{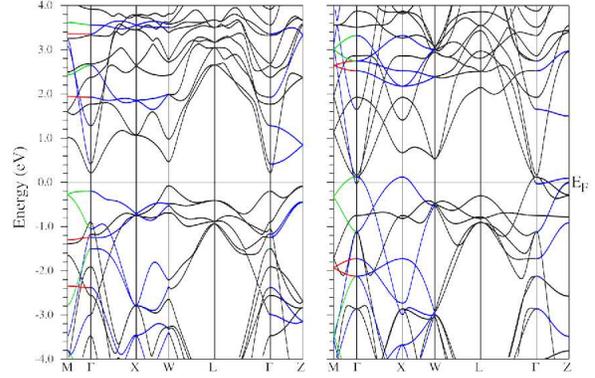}
\caption{(Color online) Band structures with YS-PBE0 ($\alpha=0.1$) for the orthorhombic AFM2 (left panel, insulating) and cubic AFM1 phase (right panel, metallic). The colors indicate the irreducible representation of the eigenvalues.}\label{bands_hybrid}
\end{figure}

As can be seen from Table \ref{table1}, the results with hybrid functionals strongly depend on the amount of HF exchange $\alpha$. If we take as a reference the values obtained within TB-mBJLDA (similar to those from LDA+$U$ with $U=4$ eV), using $\alpha=0.1$ gives closer
results than using $\alpha=0.25$. The band structure for the cubic AFM1 and orthorhombic AFM2 phase are shown in Fig. \ref{bands_hybrid} for the screened YS-PBE0 functional with
$\alpha=0.1$.

For the ground state orthorhombic AFM2 phase, there is a reduction of the band gap with respect to LDA+$U$ or TB-mBJLDA (as a consequence of the reduction of $\alpha$) to 0.2 eV. The band structures retain the same features that we discussed above for LDA+$U$ or TB-mBJLDA. In
particular, the VBM is also not at $W$, but displaced to a non-high symmetry point
(see Fig. \ref{bands_special} for LDA+$U$). Above the Fermi level, the splittings between the bands show more similarities with those obtained from LDA+$U$ with $U=4$ eV.
Figure \ref{dos} shows that the shape of the DOS is very similar to the
TB-mBJLDA DOS, but the total valence band width is increased to about 8.2 eV.
The same thing can be observed for the AFM1 phase: the band structure below the Fermi energy is more similar to the one obtained with TB-mBJLDA and above to that obtained with LDA+$U$.

Hence, hybrid functionals can also describe accurately the electronic structure of CrN, but by tuning the amount of HF exact exchange. Although results and tunability with hybrid functionals are comparable with LDA+$U$, this last method allows an easier physical description of the tuning parameter $U$ and is computationally less expensive.

\section{Discussion}

As can be seen from the values of the band gap in Table \ref{table1}, the tendency towards a band gap opening is strongest for the AFM2 ground state phase (for LDA+$U$ with $U$ as low as 3 eV, TB-mBJLDA, and hybrid functionals). Although more difficult, it is also possible to open a gap in the other AFM ordering (AFM1). 

As mentioned in Sec. \ref{intro}, the intrinsic mechanism of electronic transport in CrN is a controversial issue. Most of the discrepancies in experimental works appear among those where the properties of thin films (where by epitaxial constraints even the phase transition can be suppressed) have been measured. Hence, we will focus in the couple of works where bulk polycrystalline samples have been studied to reconcile our results with previous data.\cite{bhobe, crn_quintela_prb} 

In the work by Quintela \textit{et al.}\cite{crn_quintela_prb} the resistivity of Cr$_{1-x}$V$_x$N has been measured. Negative temperature coefficient of resistivity
($\text{TCR}=d\rho/dT$), typical for semiconducting behavior, was observed for samples with V doping $x\leq0.1$. For $x\geq0.2$ the system shows a positive TCR pointing to an itinerant electron behavior. Stoichiometric CrN exhibits $\text{TCR}<0$ (nonitinerant) over the whole $T$-range although only for $T>T_N$ the behavior can be fitted to that of an activated semiconductor with a value of the activation energy of 75 meV. In the low-temperature AFM orthorhombic phase, the mechanism of electronic transport is not fully understood: neither a simple thermal activation nor a variable range hopping behavior describe the resistivity under T$_N$. In a system with a charge gap, given the exponential dependence of $\rho(T)$ with $T$, $d(\ln\sigma)/d(\ln T)$ should diverge as $T\rightarrow0$. However, for CrN, it goes to zero as $T\rightarrow0$. Hence, despite the negative TCR, CrN cannot be classified as a typical thermally activated semiconductor. The different semiconducting behavior in the low-$T$ phase was ascribed to the changes in chemical bond linked to the AFM order. Deriving the same result for stoichiometric CrN, Bhobe \textit{et al.}\cite{bhobe} concluded that the low temperature phase of CrN is one of the rare examples of AFM-itinerant (metallic) systems in spite of negative TCR. Back to Ref. \onlinecite{crn_quintela_prb}, for $x= 0.1$ in the Cr$_{1-x}$V$_x$N series, there is a sudden change in the sign of TCR at the N\'eel temperature: $\text{TCR}>0$ (itinerant) for the high-$T$ PM phase and $\text{TCR}<0$ (nonitinerant) for the low-$T$ AFM phase. Hence, it seems that for $T<T_N$, the opening of a charge gap at the magnetic ordering temperature increases the activation energy, recovering semiconducting like behavior below $T_N$. 

The fact that the opening of a charge gap is intrinsic to the ground-state AFM ordering is consistent with our calculations. We show that the onset of magnetism should be accompanied of a gap opening. The behavior of the transport properties in the PM phase is beyond the scope of this work, but our calculations can conclude that the AFM phase is always insulating and, if it is not, then it has to do with issues caused by nanostructuring such as non-stoichiometries, strain, surface effects, etc. In a system situated close to a metal-insulator transition, small changes can induce a transition from localized to itinerant behavior, but this is not reachable in bulk-stoichiometric CrN in the magnetic phase, as our calculations show.

Further evidence of the existence of an unconventional electronic state in the AFM phase of CrN, not thermally activated neither fully itinerant, was observed in Ref. \onlinecite{crn_quintela_prb} comparing the behavior of the electrical resistivity with the thermoelectric power measurements. For stoichiometric CrN, the thermoelectric power shows linear temperature dependence (typical for metals or highly degenerate semiconductors). In addition, no discontinuity is observed at $x=0.2$, where TCR changed from negative to positive at low temperatures.  

In order to check if the experimental behavior of the thermoelectric power can be reproduced from our calculations, we have obtained its dependence with temperature in the magnetic ground state (AFM2) for LDA+$U$ ($U=4$ eV) and TB-mBJLDA calculations. This has been carried out from our band structure calculations within a semiclassical approach based on the Boltzmann transport theory through the BoltzTraP code.\cite{boltztrap} 

\begin{figure}
\includegraphics[width=\columnwidth,draft=false]{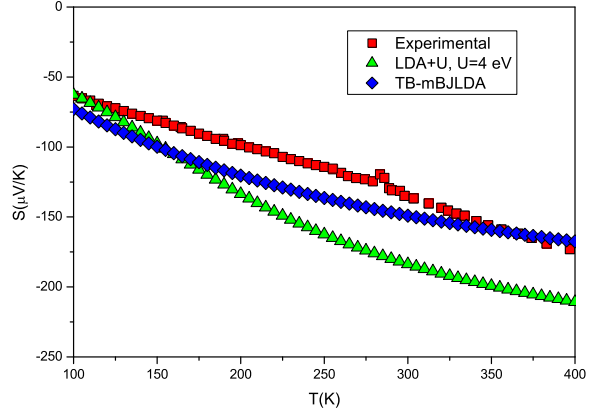}
\caption{(Color online) Experimental \cite{crn_quintela_prb} and calculated temperature dependence of the thermopower. The results for both TB-mBJLDA and LDA+$U$ ($U=4$ eV) are shown.}\label{thermopower}
\end{figure}

For the sake of comparison, we present the calculations together with the experimental values of the Seebeck coefficient taken from Ref.~\onlinecite{crn_quintela_prb} (see Fig.~\ref{thermopower}). The results for the AFM2 phase fit the experimental values
rather well, particularly in TB-mBJLDA, both in order of magnitude of the thermopower and also on the observed evolution with temperature ($dS/dT<0$) (although the behavior found from the calculations is not as linear as obtained experimentally). 

Hence, our thermopower calculations confirm that stoichiometric CrN in the magnetic phase is semiconducting, but close to a transition from localized to itinerant electronic behavior.

\section{SUMMARY AND CONCLUSIONS}

We have performed electronic structure calculations for CrN using a diverse set of exchange-correlation potentials: PBE, LDA+$U$, the semilocal functional developed by Tran and Blaha based on a modification of the Becke Johnson potential,\cite{BJ} and hybrid functionals both in screened (YS-PBE0) and unscreened (PBE0) modes.

In any case, our calculations show that the onset of magnetism in CrN should be accompanied of a gap opening. We can conclude that the AFM phase is always insulating in agreement with experiments in stoichiometric bulk samples. The different metallic behavior found in various thin films has to do with nanostructuring that could include vacancies, strain, surface effects, etc. In a system situated so close to the itinerant boundary, small changes can yield metallicity and even the suppression of the magnetic transition (which we have seen from our calculations it is strongly dependent on the structural details). The behavior obtained for the thermoelectric power confirms that stoichiometric CrN in the magnetic phase is semiconducting, but close to the transition from localized to itinerant behavior. The solution we found as ground state can describe accurately the transport properties of the system.

In addition, our results give further evidence that the TB-mBJLDA functional is very useful for treating the electronic structure of correlated semiconductors giving almost the same band gaps as obtained from LDA+$U$ with $U=4$ eV (close to the one determined experimentally) for different magnetic configurations allowing a parameter free description of the system. Hybrid functionals are also well capable of describing the electronic structure of CrN by tuning the amount of HF exact exchange. However, results and tunability are comparable with LDA+$U$, that allows an easier physical description of the tuning parameter $U$.

\acknowledgments

The authors thank the CESGA (Centro de Supercomputaci\'on de Galicia) for the computing facilities and the Ministerio de Educaci\'{o}n y Ciencia (MEC) for the financial support through the project MAT2009-08165. Authors also thank the Xunta de Galicia for the project INCITE08PXIB236053PR. A.~S. Botana thanks MEC for a FPU grant. V. Pardo thanks the Spanish Government for financial support through the Ram\'on y Cajal Program. F. Tran and P. Blaha
thank the Austrian Science Fund for financial support (Project No. SFB-F41, ViCoM).

\end{document}